\documentclass[11pt]{article}
\usepackage{amsmath, amsthm, amssymb, amsthm, dcolumn}

\usepackage{graphicx} 
\usepackage{color}
\usepackage{soul}
\usepackage{float}
\usepackage[utf8]{inputenc}
\usepackage{subcaption}
\usepackage[titletoc, title]{appendix}
\usepackage{natbib}
\usepackage{tabu,multirow}
\usepackage[titletoc]{appendix}
\usepackage{import}
\usepackage{graphicx}
\usepackage[table,xcdraw]{xcolor}
\newtheorem{definition}{Definition}

\usepackage[utf8]{inputenc}
\usepackage[flushleft]{threeparttable}
\usepackage{dcolumn}
\usepackage{caption}
\usepackage{soul}
\usepackage{ragged2e}
\usepackage{amsmath}

\begin{document}

\begin{titlepage}
	\begin{center}
	{\huge Detecting common bubbles in multivariate mixed causal-noncausal models\par }
	\vspace{1cm}
	\renewcommand{\thefootnote}{\alph{footnote}} 
	{\Large Gianluca Cubadda\footnote{Tor Vergata University of Rome School of Economics}, Alain Hecq\footnote{Maastricht University School of Business and Economics}\footnote{Corresponding author : Alain Hecq, Maastricht University, Department of Quantitative Economics, School of Business and Economics, P.O.box 616, 6200 MD, Maastricht, The Netherlands. Email: a.hecq@maastrichtuniversity.nl.\par}  and Elisa Voisin\footnotemark[2]
	}\\
	\vspace{0.2cm}
	{\large Maastricht University}

	\vspace{1cm}
	{\large June, 2022\par}
	
	\vspace{1cm}
	\end{center}
\begin{abstract}
    This paper proposes methods to investigate whether the bubble patterns observed in individual series are common to various series. We detect the non-linear dynamics using the recent mixed causal and noncausal models. Both a likelihood ratio test and information criteria are investigated, the former having better performances in our Monte Carlo simulations. Implementing our approach on three commodity prices we do not find evidence of commonalities although some series look very similar.
\end{abstract}

	\vspace{0.5cm}
	\justifying

	\textbf{Keywords:} Forward-looking models, bubbles, co-movements \\
	\textbf{JEL.} C32
	
\end{titlepage}

\graphicspath{ {Pictures/} }

\section{Introduction}
Economic and financial time series exhibit many distinctive characteristics among which the presence of serial correlation, some seasonality, stochastic or deterministic trends, time varying volatility, non-linearities. However, in multivariate analyses, namely when one investigates relationships between variables, it is frequent to observe that one or more of these features that were detected in individual series are common to several variables. We talk about common features when such features are annihilated with some suitable combinations. The most famous example is probably cointegration, that is the presence of common stochastic trends \citep{engle1987co}. Other forms of co-movements \citep{engle1993testing} have also been studied, giving rise to developments around the notions of common cyclical features \citep{vahid1993common}, common deterministic seasonality \citep{engle1996common}, common volatility \citep{engle1993common}, co-breaking \citep{hendry2007co}, etc. Recognizing these common feature structures presents numerous advantages from an economic perspective (e.g. the whole literature on the existence of long-run relationships). There are also several implications for statistical modeling. Imposing the commonalities helps to reduce the number of parameters that must be estimated. That potentially leads to efficiency gains and to improvements in forecasts accuracy \citep{issler2001common}. The factor structure underlying common features can also be used to forecast a set of time series using only the forecasts of the common component and the estimated loadings. \\

Building on such a common features approach, we propose in this paper to detect the presence of common bubbles in stationary time series. Intuitively, the idea is to detect in univariate time series bubble patterns and to investigate whether those bubbles would be common to a set of assets. In the affirmative, a portfolio composed of those series would not have such a non-linear local explosive characteristic. There are several ways to capture bubbles in the data. We rely on mixed causal-noncausal models (denoted MAR$(r,s)$ hereafter), namely autoregressive time series that depend on both \textit{r} lags and \textit{s} leads. There is indeed a recent interest in the properties of noncausal processes associated with a blooming of applications on commodity prices, inflation or cryptocurrency series as well as the developments around the notion of non-fundamental shocks. See \citet{voisin2021oil} for the references therein. We choose to consider mixed causal and noncausal models as they might also be used for forecasting. This is not necessarily the case with other approaches aiming at identifying bubble phases. In a MAR($r,s$) framework, in the presence of common bubbles among a set of time series, several variables can be forecasted with one of the series (and their loadings).\\

\citet{cubadda2019detecting} extend the canonical correlation framework of \citet{vahid1993common} from purely causal vector autoregressive models (namely the traditional serial correlation common feature approach within a VAR) to purely noncausal VARs (a VAR with leads only). They show that more commonalities emerge when we also look at VARs in reverse time. The tests statistics they developed do not generally work for mixed models though. Consequently, we extend their work and we propose a likelihood ratio test that compares the unrestricted multivariate vector mixed causal-noncausal model (hereafter VMAR$(r,s)$) with its restricted version in which reduced ranks are imposed on the lead polynomial matrix, which is our notion of common bubbles. Indeed, this is equivalent to require that there exists a linear combination of variables with bubbles that does not possess the bubble feature. See for instance \citet{cubadda2021reduced} for a recent survey on reduced rank techniques for common features. We also consider the use of information criteria as an alternative strategy. \\

The rest of this paper is as follows: in Section \ref{VMAR-sec:Model} we set up the notations for multivariate mixed causal and noncausal models. Contrarily to the univariate case, two distinct multivariate multiplicative representations lead to the same linearized form. They consequently also have the same likelihood but with different lag-lead polynomial matrices. We advocate to use the VMAR with the lead polynomial first as the alternative matrix polynomial structure does not allow to discover the presence of common bubbles even when they are present. We explain how to implement the likelihood ratio test that we introduce in this paper. Section \ref{VMAR-sec:MC} investigates, using Monte Carlo simulations, the small sample properties of our strategy for a bivariate and three dimensional systems both under the null of common bubbles and the alternative of no rank reductions. Section \ref{VMAR-sec:empirical} illustrates our approach on three commodity prices. The presence of common bubbles is rejected in every bivariate and three dimensional systems although from the graphs, series looked rather similar. Section \ref{VMAR-sec:conclusion} concludes.

\section{Multivariate mixed causal-noncausal models}\label{VMAR-sec:Model}
Recall that a univariate MAR($r,s$) model is constructed as follows,
\begin{equation*}
    (1-\phi_1L-\hdots-\phi_rL^r)(1-\psi_1L^{-1}-\hdots-\psi_sL^{-s})y_t=e_t,
\end{equation*}
where all coefficients are scalars and thus the model is commutative, 
\begin{equation*}
    (1-\psi_1L^{-1}-\hdots-\psi_sL^{-s})(1-\phi_1L-\hdots-\phi_rL^r)y_t=e_t.
\end{equation*}
That is, either of the two representations will yield the same coefficients. $L^r$ is the lag operator such that $L^ry_t=y_{t-r}$ and $L^{-s}$ is the lead operator such that $L^{-s}y_t=y_{t+s}$.\\

Let us now consider $Y_{t}$, an $N$ dimensional stationary process. We assume for notation simplicity that deterministic elements such as the intercept or seasonal dummies have been subtracted. Analogously to the univariate case, a multivariate mixed causal-noncausal model with \textit{r} lags and \textit{s} leads, denoted VMAR$(r,s)$, is defined in its multiplicative forms as follows,
\begin{equation}
\Psi (L^{-1})\Phi (L)Y_{t}=\varepsilon _{t},
\label{VMAR-eq:VMAR_multiplicative_psi_phi}
\end{equation}%
\begin{equation}
\bar{\Phi}(L)\bar{\Psi}(L^{-1})Y_{t}=\bar{\varepsilon}_{t}.
\label{VMAR-eq:VMAR_multiplicative_phi_psi}
\end{equation}
where, 
\begin{align*}
    \Psi(L^{-1})\Phi(L)&=(I_N-\Psi_1L^{-1}-\hdots-\Psi_sL^{-s})(I_N-\Phi_1L^{1}-\hdots-\Phi_s L^{r})\\
    \bar{\Psi}(L^{-1})\bar{\Psi}(L)&=(I_N-\bar{\Psi}_1L^{-1}-\hdots-\bar{\Psi}_sL^{-s})(I_N-\bar{\Psi}_1L^{1}-\hdots-\bar{\Psi}_s L^{r}).
\end{align*}

To simplify the analyses, ${\varepsilon }_{t}$ and $\bar{\varepsilon}_{t}$ follow multivariate Student's \textit{t}-distributions. We can consider other distributions than the Student as long as they are non Gaussian. This is indeed the condition that allows for the distinction of the causal, noncausal or mixed specifications. Both models \eqref{VMAR-eq:VMAR_multiplicative_psi_phi} and \eqref{VMAR-eq:VMAR_multiplicative_phi_psi} are equivalent but contrary to univariate MAR models, they are two distinct representations of the same process, given the non commutativity property of the matrix product. This means that the lag polynomial matrices ${\Phi }(L)$ and $\bar{\Phi}(L),$ though of the same order \textit{r}, have different values for coefficient matrices. The same observation applies to the lead polynomials ${\Psi }(L^{-1})$ and $\bar{\Psi}(L^{-1})$ which are of the same order \textit{s}. We assume that the roots of the determinants of each polynomial matrices $\Psi(L^{-1}),\Phi(L)$, $\bar{\Phi}(L),\bar{\Psi}(L^{-1})$ are outside the unit circle to fulfill the stationarity condition. Furthermore, we will show later that the distribution of the errors ${\varepsilon }_{t}$ and $\bar{\varepsilon}$ have identical degrees of freedom $\lambda $ but different scale matrices. We denote them by $\Sigma $ and $\bar{\Sigma}$ respectively, both being symmetric positive definite matrices. \\

Let us further denote $A(L)$ and $\bar{A}(L)$ the expanded products\footnote{%
This is the restricted linear form that is used in the maximum likelihood
estimation. \citet{gourieroux2017noncausal} have proposed an alternative
approach based on roots inside and outside the unit circle of an
autoregressive polynomial.} of the lag and lead polynomials in the two models with 
\begin{align*}
\Psi (L^{-1})\Phi (L)& \equiv A(L)=\sum_{j=-s}^{r}A_{j}L^{j}\quad
\rightarrow \quad A(L)Y_{t}=\varepsilon _{t}, \\
\bar{\Phi}(L)\bar{\Psi}(L^{-1})& \equiv \bar{A}(L)=\sum_{j=-s}^{r}\bar{A}%
_{j}L^{j}\quad \rightarrow \quad \bar{A}(L)Y_{t}=\bar{\varepsilon}_{t}.
\end{align*}%
The general forms of the expansion of the lead and lag polynomials for each
representation are {\small 
\begin{equation}
    \begin{split}
        A& (L)\equiv \sum_{j=-s}^{r}A_{j}L^{j} = \\
        & I+\sum_{i=1}^{\substack{min\\ \{r,s\}}}\Psi _{i}\Phi_{i}-\sum_{i=1}^{r}\left( \Phi _{i}-\sum_{\substack{\forall \{l,m\} \\s.t. \\l-k=i}}\Psi _{l}\Phi _{m}\right) L^{i}-\sum_{j=1}^{s}\left(\Psi _{j}-\sum_{\substack{\forall \{l,m\} \\ s.t. \\m-l=j}}\Psi_{l}\Phi _{m}\right) L^{-j}, \\
        \bar{A}& (L)\equiv\sum_{j=-s}^{r}\bar{A}_{j}L^{j}= \\
        & I+\sum_{i=1}^{\substack{min\\ \{r,s\}}}\bar{\Phi}_{i}\bar{\Psi}_{i}-\sum_{i=1}^{r}\left(\bar{\Phi}_{i}-\sum _{\substack{\forall \{l,m\}\\ s.t. \\m-l=i}}\bar{\Phi}_{m}\bar{\Psi}_{l}\right)L^{i}-\sum_{j=1}^{s}\left(\bar{\Psi}_{j}-\sum_{\substack{\forall\{l,m\} \\ s.t. \\m-l=j}}\bar{\Phi}_{m}\bar{\Psi}_{l}\right) L^{-j},
    \end{split}
    \label{VMAR-eq:expanded}
\end{equation}

}with $1\leq l\leq s$ and $1\leq m\leq r$. This shows that both multiplicative representations yield the exact same additive form, 
\begin{equation}
\underset{\substack{A_0^{-1}A(L)\\=\\\bar{A}_0^{-1}\bar{A}(L)}}{
\underbrace{B\text{{(}}L\text{{)}}}}Y_{t}=\underset{\substack{A_0^{-1}\varepsilon_t\\=
\\ \bar{A}_0^{-1}\bar{\varepsilon}_t}}{\underbrace{\eta _{t}}},
\label{VMAR-eq:additive}
\end{equation}%

where $\eta _{t}$ follows a multivariate Student-\textit{t} distribution with degrees of freedom $\lambda $ -- analogously to $\varepsilon _{t}$ and $\bar{\varepsilon}_{t}$ from representations \eqref{VMAR-eq:VMAR_multiplicative_psi_phi} and \eqref{VMAR-eq:VMAR_multiplicative_phi_psi} -- and with a scale matrix $\Omega=A_0^{-1}\Sigma (A_0^{-1})^{^{\prime }}=\bar{A}_0^{-1}\bar{\Sigma}(\bar{A}_0^{-1})^{^{\prime }}$. The lag polynomial in \eqref{VMAR-eq:additive} is the following, 
\begin{equation}\label{VMAR-eq:B(L)}
    B(L)=I-\sum_{i=1}^{r}B_{i}L^{i}-\sum_{j=1}^{s}B_{-j}L^{-j}.
\end{equation}
The example of derivations of the coefficients of a VMAR(2,2) is given in Section \ref{VMAR-subsec:CB_rs}. \\

Contrary to the univariate case, with commutative
multiplicative form and therefore a unique solution, a multivariate VMAR$(r,s)$ processes has two distinct representations. The multivariate process can be estimated with either of the multiplicative representations \eqref{VMAR-eq:VMAR_multiplicative_psi_phi} and \eqref{VMAR-eq:VMAR_multiplicative_phi_psi}. While the coefficient matrices will differ, both representations will correspond to the same expanded form of the model \eqref{VMAR-eq:additive}. This makes however the interpretation of the lag and lead coefficient matrices in the multiplicative forms more intricate. \citet{lanne2013noncausal} advocate for the use of one or the other representation depending on the analysis performed; one representation might be easier to employ for certain inquiries.

\subsection{Common bubbles in VMAR(\textit{r},\textit{s})}\label{VMAR-subsec:CB_rs}

Now that we have set up the notations of the unrestricted multivariate mixed model we consider additional restrictions coming from commonalities in the lead polynomial matrix. Indeed it is the lead component that induces some non-linearities similar to the bubble pattern \citep{gourieroux2013explosive}.Although the focus in this paper is on common bubbles, our approach can be easily extended the investigation to commonalities in the causal part or in both the lag and the lead components.

\begin{definition}
The $N$ dimensional process $Y_{t}$ displays common bubbles (hereafter CB) if there exists a matrix $\delta$ of dimension $N\times k$, with $0<k<N$, such that, $\delta ^{\prime }B_{-j}=0$ for $j=1,\dots ,s$, where the coefficient matrix $B_{-j}$ is a matrix of the expanded lag polynomial \eqref{VMAR-eq:B(L)}. This implies that the coefficient matrices $B_{-j}$ can be decomposed as $\delta_{\bot }\beta _{j}^{\prime }$ where $\delta _{\bot }$ is the $N\times (N-k)$ orthogonal complement of $\delta ^{\prime }$ such that $\delta ^{\prime}\delta _{\bot }=0$ and $\beta _{j}^{\prime }$ is a matrix with dimension $(N-k)\times N$. 
\end{definition}

Let us start from the example $r=s=2$. The coefficient matrices of the leads in the additive representation \eqref{VMAR-eq:additive} with reduced rank restrictions are%
\begin{align*}
B_{-1}& =A_0^{-1}(\Psi _{1}-\Psi _{2}\Phi _{1}) & B_{-2}& =A_0^{-1}\Psi
_{2} \\
& =\bar{A}_0^{-1}(\bar{\Psi}_{1}-\bar{\Phi}_{1}\bar{\Psi}_{2}) & & =\bar{A}_0^{-1}\bar{\Psi}_{2} \\
& =\delta _{\bot }\beta _{1}^{\prime }, & & =\delta _{\bot }\beta
_{2}^{\prime },
\end{align*}%
where the matrices $A_0$ and $\bar{A}_0$ have been derived from the expanded lag polynomials \eqref{VMAR-eq:expanded} with 
\begin{align*}
A_0& =(I_{N}+\Psi _{1}\Phi _{1}+\Psi _{2}\Phi _{2}) \\
\bar{A}_0& =(I_{N}+\bar{\Phi}_{1}\bar{\Psi}_{1}+\bar{\Phi}_{2}\bar{\Psi}%
_{2}).
\end{align*}%
Hence, the matrix $\delta ^{\prime }$ of dimension $k\times N$, with $0<k<N$ annihilates the forward looking dynamics 
\begin{equation*}
\delta ^{\prime }B_{-1}=\delta ^{\prime }B_{-2}=0.
\end{equation*}%
This implies that, for the second lead coefficients, 
\begin{equation*}
\delta ^{\prime }B_{-2}=\delta ^{\prime }A_0^{-1}\Psi _{2}=\delta ^{\prime }%
\bar{A}_0^{-1}\bar{\Psi}_{2}=0.
\end{equation*}%
Since $\delta ^{\prime }A_0^{-1}$  cannot be equal to zero, it implies that $\delta ^{\prime}A_0^{-1}=\gamma ^{\prime }$ (resp. $\delta ^{\prime }\bar{A}_0^{-1}=\bar{\gamma}^{\prime }$) , where $\gamma$ (resp. $\bar{\gamma}$) is some $N\times k$ dimensional matrix and thus $\gamma ^{\prime }\Psi_{2}=0$ (resp. $\bar{\gamma}^{\prime }\bar{\Psi}_{2}=0$). Hence, both $\Psi_{2}$ and $\bar{\Psi}_{2}$ must have rank $N-k$, but potentially different left null spaces \citep[see also][]{cubadda2019detecting}. \\

For the first lead coefficient of representation %
\eqref{VMAR-eq:VMAR_multiplicative_psi_phi}, 
\begin{align*}
\delta ^{\prime }B_{-1}& =\delta ^{\prime }A_0^{-1}(\Psi _{1}-\Psi _{2}\Phi_{1}) \\
& =\gamma ^{\prime }(\Psi _{1}-\Psi _{2}\Phi _{1}) \\
& =\gamma ^{\prime }\Psi _{1}=0
\end{align*}%
which implies that $\Psi _{1}$ and $\Psi _{2}$ must have the same left null space. For the alternative representation in \eqref{VMAR-eq:VMAR_multiplicative_phi_psi},
\begin{align*}
\delta ^{\prime }B_{-1}& =\delta ^{\prime }\bar{A}_0^{-1}(\bar{\Psi}_{1}-%
\bar{\Phi}_{1}\bar{\Psi}_{2}) \\
& =\bar{\gamma}^{\prime }(\bar{\Psi}_{1}-\bar{\Phi}_{1}\bar{\Psi}_{2})=0
\end{align*}%
which implies that $\bar{\gamma}^{\prime }\bar{\Psi}_{1}=\bar{\gamma}^{\prime }\bar{\Phi}_{1}\bar{\Psi}_{2}$ and that $\bar{\Psi}_{1}$ might not necessarily even be a reduced-rank matrix.\\

Since $\Psi_1$ and $\Psi_2$ of representation \eqref{VMAR-eq:VMAR_multiplicative_psi_phi} must have the same left null space and keeping in mind that $\delta'A_0^{-1}=\gamma'$ we have, 
\begin{align*}
\delta^{\prime }&=\gamma'A_0= \gamma^{\prime }(I_{N}+\Psi_{1}\Phi_{1}+\Psi_{2}\Phi_{2})= \gamma^{\prime },
\end{align*}
which shows that $\Psi_1$ and $\Psi_2$ have the same left null space as $B_{-1}$ and $B_{-2}$. \\

For representation \eqref{VMAR-eq:VMAR_multiplicative_phi_psi}, only the second lead coefficient matrix must have reduced rank. The lead coefficients of the two representations in the presence of CB are the following, 
\begin{align*}
\Psi_1 &= A_0\delta_\bot\big(\beta^{\prime }_1 + \beta^{\prime }_2\Phi_1%
\big) = (I_{N}+\Psi_{1}\Phi_{1}+\Psi_{2}\Phi_{2})\delta_\bot\big(%
\beta^{\prime }_1 + \beta^{\prime }_2\Phi_1\big), \\
\Psi_2 &= A_0\delta_\bot \beta^{\prime }_2 =
(I_{N}+\Psi_{1}\Phi_{1}+\Psi_{2}\Phi_{2})\delta_\bot \beta^{\prime }_2, \\ \\
\bar{\Psi}_1 &= \bar{A}_0\delta_\bot\beta^{\prime }_1+ \bar{\Phi}_1\bar{A}_0\delta_\bot\beta^{\prime }_2 \\
&= (I_{N}+\bar{\Phi}_{1}\bar{\Psi}_{1}+\bar{\Phi}_{2}\bar{\Psi}%
_{2})\delta_\bot\beta^{\prime }_1+ \bar{\Phi}_1(I_{N}+\bar{\Phi}_{1}\bar{\Psi%
}_{1}+\bar{\Phi}_{2}\bar{\Psi}_{2})\delta_\bot\beta^{\prime }_2,\\
\bar{\Psi}_{2} &= \bar{A}_0\delta_\bot \beta^{\prime }_2 = (I_{N}+\bar{\Phi}%
_{1}\bar{\Psi}_{1}+\bar{\Phi}_{2}\bar{\Psi}_{2})\delta_\bot \beta^{\prime
}_2.
\end{align*}
While it is clear that $\Psi_1$, $\Psi_2$, $B_{-1}$ and $B_{-2}$, having the same left null space, satisfies the condition for the presence of a CB, we can see that pre-multiplying $\bar{\Psi}_1$ and $\bar{\Psi}_2$ by $\delta'$ will not annihilate the dynamics. This due the matrix multiplication structure of their components, as revealed by the right-hand sides of the equations. For instance, from the equation of $\bar{\Psi}_2$,$\delta'(I_{N}+\bar{\Phi}_{1}\bar{\Psi}_{1}+\bar{\Phi}_{2}\bar{\Psi}_{2})$ would have to simplify to $\delta'$ to annihilate the dynamics, however, this would require restrictions on the lag coefficients.\\

Overall, it is easy to see that the same conclusion applies for any VMAR(\textit{r},\textit{s}). In the presence of CB, all lead matrices of representation \eqref{VMAR-eq:VMAR_multiplicative_psi_phi} must have the same left null space. Hence, for the investigation of CB this representation is the most appropriate and straightforward to allow for tests. 

\subsection{Testing for common bubbles}

In an $\mathit{N-}$dimensional VMAR$(\textit{r},\textit{s})$, our definition of common bubbles implies that there exists a full-rank $N\times k$ matrix $\delta$, with $0<k<N$ that annihilates the noncausal dynamics of the multivariate process \eqref{VMAR-eq:additive}, i.e. $\delta ^{\prime }B_{-j}=0$ for all $j=1,\dots ,s$. It therefore entails that $\delta ^{\prime }Y_{t}$ is a purely causal $k$-dimensional process. We showed above that in the presence of CB all lead matrices estimated from the multiplicative representation \eqref{VMAR-eq:VMAR_multiplicative_psi_phi} must have the same left null space as the coefficients $B_{-j}$. They can thus be decomposed as 
\begin{equation}\label{VMAR-eq:decomposition_psi}
\Psi _{j}=\delta _{\bot }\Gamma _{j}^{\prime },\quad \quad \text{with}\quad
j=1,\dots ,s,
\end{equation}
with $\delta _{\bot }$ the aforementioned $N\times (N-k)$ matrix such that $%
\delta ^{\prime }\delta _{\bot }=0$ and $\Gamma _{j}^{\prime }$ a $(N-k)\times N$ full-rank matrix. \\

We suggest a likelihood ratio test comparing the likelihood value of the unrestricted model \eqref{VMAR-eq:VMAR_multiplicative_psi_phi} with the likelihood value of the restricted model 
\begin{equation}  \label{VMAR-eq:VMAR_restricted}
(I_N -\delta_{\bot}\Gamma^{\prime }_1 L^{-1} -\dots-
\delta_{\bot}\Gamma^{\prime }_s L^{-s})(I_N - \Phi_1 L -\dots- \Phi_r
L^r)Y_t =\varepsilon_t,
\end{equation}
where the coefficient matrices are as defined in \eqref{VMAR-eq:decomposition_psi}. Since $\delta_\bot$ has dimension $N\times (N-k)$ with $0<k<N$, there are $N-1$ possible reduced-rank model to consider, for all possible $k$. Furthermore, the matrix $\delta^{\prime}$ is normalized, such that $\delta'=\left [I_k, \; \delta^*\right]$ and thus only involves $k\times (N-k)$ free parameters in $\delta^*$. Hence, the test statistic based on the log-likelihood values, 
\begin{equation}  \label{VMAR-eq:LR}
    LR_k=-2\big[\text{ln}({L}_0^{(k)})-\text{ln}(\hat{L})\big],
\end{equation}
follows, under the null of CB, a $\chi^2_\rho$ distribution, with $\rho=k^2-Nk(1-s)$, against the alternative of a full rank. \\

An alternative is to employ information criteria as a selection method between the restricted and unrestricted specifications, 
\begin{equation}  \label{VMAR-eq:BIC}
    BIC_k= K \,\text{ln}(T)-2\,\text{ln}(\hat{L}),
\end{equation}
\begin{equation}  \label{VMAR-eq:AIC}
    AIC_k= 2K - 2\,\text{ln}(\hat{L}),
\end{equation}
with \textit{K} the number of coefficients estimated in the lag and lead matrices of the model -- for the unrestricted model, $K_u=N^2(r+s)$ and for the restricted models, $K_{r_k}=K_u-\big(k^2-Nk(1-s)\big)=K_u-\rho$.\\

Alternatively, one can test the null hypothesis that the lead coefficient matrices have rank $0<(N-k)<N$ against the alternative that they do have commonalities, however that they have a larger rank $(N-k)<(N-l)<N$. In such case, the difference in the number of estimated coefficients $\rho= N(1-s)(k-l)+l^2-k^2$ is smaller than against the alternative of full rank.

\section{Monte Carlo analysis}\label{VMAR-sec:MC}
We investigate using Monte Carlo simulations the performance of our strategies to detect common bubbles in bivariate and trivariate VMAR(1,1) models. We consider two sample sizes ($T=500$ and $1000)$ and two different degrees of freedom of the error term with very leptokurtic distributions, namely $\lambda=3$ and 1.5, to respectively consider a finite and infinite variance case. We employ lead coefficient matrices with and without reduced rank to analyse the detection of the correct model under the null of common bubbles and under the alternative of no such co-movements. The coefficients employed in the bivariate settings are displayed in Table \ref{VMAR-tab:param_MC_N2}. \\

\begin{table}[h!]
\caption{Monte Carlo parameters for bivariate VMAR(1,1)}
\label{VMAR-tab:param_MC_N2}\centering
\begin{tabular}{lll}
\hline
 \multirow{3}{*}{$\Phi=\begin{bmatrix} 0.5 & 0.1 \\0.2 & 0.3 \end{bmatrix}$}  & \multirow{3}{*}{}  & \multirow{3}{*}{$\Sigma=\begin{bmatrix}4 & 0.5 \\0.5 & 1\end{bmatrix}$}  \\ \\ \\
 \multirow{2}{*}{$T=\big\{500,\;1\,000\big\}$} &\multirow{2}{*}{}&\multirow{2}{*}{} \\ \\
 \multirow{2}{*}{$\lambda=\big\{1.5,\;3\big\}$} & \multirow{2}{*}{} &\multirow{2}{*}{}   \\ \\
\multicolumn{3}{l}{\multirow{7}{*}{
                                    $\Psi=\begin{cases} 
                                            \begin{bmatrix}0.3 & 0.25 \\0.6 & 0.5 \end{bmatrix}
                                            =  \begin{bmatrix} 1 \\2\end{bmatrix}
                                            \begin{bmatrix}0.3 & 0.25\end{bmatrix}
                                            & (H_0: \,\text{CB}) \\ \\ 
                                            \begin{bmatrix} 0.1 & 0.4 \\0.6 & 0.5 \end{bmatrix} 
                                            & (H_1: \,\text{no CB}) 
                                    \end{cases}$}}  \\ \\ \\ \\ \\ \\ \\ \hline
\end{tabular}
\end{table}

Results, based on 3\thinspace 000 replications for each combination of parameters, are reported in Table \ref{VMAR-tab:MC_results_N3}.\footnote{Estimating multivariate causal-noncausal models using maximum likelihood presents high sensitivity to starting values. We do not investigate this matter here and therefore employ true values as starting values in the estimations.} All entries are the frequency of correctly detected model. That is, under the null of a CB, we report the proportion of correctly detected CB, and under the alternative of no CB, we report the proportion of correctly rejected CB. We hence perform the test $H_{0}:rank(\Psi )=1$ against the alternative that the rank is 2. The LR tests are performed at a 95\% confidence level. The information criteria detect a CB when the IC of the restricted model is lower than the one of the unrestricted model. \\

\begin{table}[h!]
\caption{MC results for N=2}
\label{VMAR-tab:MC_results_N2}\centering
\resizebox{\textwidth}{!}{%
\begin{tabular}{llccccccc}
\hline
&  & \multicolumn{7}{c}{$\lambda=3$} \\ \cline{3-9}
&  & \multicolumn{3}{c}{T=500} &  & \multicolumn{3}{c}{T=1000} \\ 
\cline{3-5}\cline{7-9}
 DGP&  & LR test & BIC & AIC &  & LR test & BIC & AIC \\ \hline
With CB (rank 1) &  & 0.946 & 0.989 & 0.838 &  & 0.944 & 0.993 & 0.834 \\ 
Without CB (rank 2) &  & 0.999 & 0.994 & 1.000 &  & 1.000 & 1.000 & 1.000 \\ \hline
&  &  &  &  &  &  &  &  \\ 
&  & \multicolumn{7}{c}{$\lambda=1.5$} \\ \cline{3-9}
&  & \multicolumn{3}{c}{T=500} &  & \multicolumn{3}{c}{T=1000} \\ 
\cline{3-5}\cline{7-9}
DGP&  & LR test & BIC & AIC &  & LR test & BIC & AIC \\ \hline
With CB (rank 1) &  & 0.913 & 0.968 & 0.779 &  & 0.914 & 0.977 & 0.783 \\ 
Without CB (rank 2) &  & 0.999 & 0.999 & 0.999 &  & 1.000 & 1.000 & 1.000 \\ \hline
\end{tabular}%
}
\\
\justifying {\footnotesize \ Based on 3000 iterations. All results are the frequencies of correctly detected models. The LR test is performed at a 95\% confidence level. For the IC, the favoured model is the one with the lowest IC value. The ranks refer to the rank of the lead coefficient in the DGP.}
\end{table}

We can notice that the frequency of Type I errors of the LR test increases when the variance of the errors becomes infinite. It however does not significantly decrease when the sample size gets larger. With finite variance ($\lambda=3$) the LR test has an appropriate size of around 5.5\% and it increases to around 8.6\% when the degrees of freedom of the errors distribution reach 1.5. Under the alternative, the LR test has a power of at least 99.9\% across all parameters combinations implying that it almost never detects a CB when there are none. \\

Regarding the model selection using information criteria, results show that BIC outperforms AIC. Under the null of a CB, BIC selects the correct model specification in 98.9\% of the cases with finite variance and a sample size of 500. The frequency increases to  99.3\% when the sample size increases to 1\,000. AIC on the other hand selects the correct model in only 83.8\% of the cases and does not increase with the sample size. The frequency of correctly selected model decreases for both when in the infinite variance case, but more drastically for AIC, which decreases to around 78\%. BIC still selects the correct model for 96.8\% of the cases with a sample size $T=500$, and the frequency increases to 97.7\% for $T=1\,000$. Under the alternative of no CB however, both IC correctly select the unrestricted specification in more than 99.4\% across all parameters combinations. \\

We now turn to the trivariate case. Now, in the presence of a CB, the rank of the lead coefficient matrix can be either 1 or 2. We thus consider the two possible CB structures. The parameters of the data generating processes are displayed in Table \ref{VMAR-tab:MC_results_N3}. \\

\begin{table}[h]
\caption{Monte Carlo parameters for trivariate VMAR(1,1)}
\label{VMAR-tab:param_MC_N3}\centering
\resizebox{\textwidth}{!}{\begin{tabular}{lllll}
\hline
\multirow{4}{*}{$\Phi=\begin{bmatrix} 0.5 & 0.1 & 0.2 \\0.2 & 0.3 & 0.1 \\0.1 & 0.4 & 0.6 \end{bmatrix}$} 
                                    & \multirow{4}{*}{}  
                                    & \multirow{4}{*}{\quad\quad\quad$\Sigma=\begin{bmatrix} 2 & 0.5 & 0.5 \\0.5 & 1 & 0.5 \\0.5 & 0.5 & 4\end{bmatrix}$}  \\ \\ \\ \\
\multirow{2}{*}{$T=\big\{500,\;1\,000\big\}$} &\multirow{2}{*}{}&\multirow{2}{*}{} \\ \\
\multirow{2}{*}{$\lambda=\big\{1.5,\;3\big\}$} 
                                        & \multirow{2}{*}{} &\multirow{2}{*}{}   \\ \\
\multicolumn{3}{l}{\multirow{7}{*}{
                                    $\Psi=
                                        \begin{cases} 
                                            \begin{bmatrix}0.3 & 0.1  & 0.1 \\0.2 & 0.3 & 0.4 \\0.7 & 0.35 & 0.4\end{bmatrix}
                                            =  \begin{bmatrix}1 & 0 \\0 & 1 \\2 & 0.5\end{bmatrix}
                                            \begin{bmatrix}0.3 & 0.1 & 0.1 \\0.2 & 0.3 & 0.4\end{bmatrix}
                                            & (H_0:\, \text{1 CB feature}) \\ \\ 
                                            \begin{bmatrix}0.15 & 0.25  & 0.4\\ 0.3 & 0.5 & 0.8 \\0.075 & 0.125 & 0.2\end{bmatrix} 
                                            =  \begin{bmatrix}1 \\2 \\0.5\end{bmatrix}
                                            \begin{bmatrix}0.15 & 0.25 & 0.4\end{bmatrix}
                                            & (H_0:\, \text{2 CB features}) \\ \\  
                                            \begin{bmatrix}0.3 & 0.2 & 0.1 \\0.2 & 0.5 & 0.4 \\0.7 & 0.125 & 0.2\end{bmatrix} &(H_1:\, \text{no CB feature})
                                        \end{cases}$}}  \\ \\ \\ \\ \\ \\ \\ \\ \\ \\ \\ \\ \\ \\ \hline
\end{tabular}
}
\end{table}

We evaluate our approach with 1\,500 replications with each of the parameters combinations. Under the null of a CB we test the correct CB specification against the alternative of the unrestricted full rank model. Under the alternative of no CB we test for each of the CB specifications.\footnote{Results for other tests, such as 1 vs 2 when the true rank is 2 for instance, are available upon requests.} Table \ref{VMAR-tab:MC_results_N3} reports the frequencies of correctly detected models either with the LR test or with model selection using the information criteria. Analogously to the bivariate case, the LR tests are performed at a 95\% confidence level and the information criteria detect a CB when the IC of the restricted model is lower than the one of the unrestricted model. \\

\begin{table}[h!]
\caption{MC results for N=3}
\label{VMAR-tab:MC_results_N3}\centering
\resizebox{\textwidth}{!}{
\begin{tabular}{cclccccccc} \hline
                                            &             &  & \multicolumn{7}{c}{$\lambda=3$}                           \\\cline{4-10}
                                            &             &  & \multicolumn{3}{c}{T=500} &  & \multicolumn{3}{c}{T=1000} \\ \cline{4-6} \cline{8-10}
$rank(\Psi)$                                           &      Rank test        &  & LR     & BIC     & AIC    &  & LR     & BIC     & AIC     \\ \hline
2 & 2 vs 3 &  & 0.944 & 0.984 & 0.817 &      & 0.951 & 0.992 & 0.843 \\
                                            &             &  &        &         &        &  &        &         &         \\ 
1                    & 1 vs 3 &  & 0.919 & 1.000 & 0.871 &      & 0.933 & 1.000 & 0.883 \\
                                            &             &  &        &         &        &  &        &         &  \\
\multirow{2}{*}{3}                                       & 2 vs 3 &  & 0.695 & 0.481 & 0.855 &      & 0.932 & 0.802 & 0.970 \\
                                            & 1 vs 3 &  & 1.000 & 1.000 & 1.000 &      & 1.000 & 1.000 & 1.000 \\
\hline \\
                                            &             &  & \multicolumn{7}{c}{$\lambda=1.5$}                           \\\cline{4-10}
                                            &             &  & \multicolumn{3}{c}{T=500} &  & \multicolumn{3}{c}{T=1000} \\\cline{4-6} \cline{8-10}
 $rank(\Psi)$                                           &     Rank test         &  & LR     & BIC     & AIC    &  & LR     & BIC     & AIC     \\ \hline
2                                      &2  vs 3 &  & 0.915 & 0.972 & 0.775 &      & 0.907 & 0.978 & 0.776 \\
                                            &             &  &        &         &        &  &        &         &         \\
1                     & 1 vs 3 &  & 0.857 & 0.998 & 0.774 &      & 0.860 & 0.999 & 0.783 \\
                                            &             &  &        &         &        &  &        &         &         \\
                                            
\multirow{2}{*}{3}                                       & 2 vs 3 &  &0.997 & 0.994 & 0.999 &      & 1.000 & 1.000 & 1.000\\
                                            & 1 vs 3 &  &1.000 & 1.000 & 1.000 &      & 1.000 & 1.000 & 1.000\\
\hline      
\end{tabular} 
}
\footnotesize \justifying
       Based on 1\,500 iterations. All results are the frequencies of correctly detected models. The LR test is performed at a 95\% confidence level. For the IC, the favoured model is the one with the lowest IC value. The ranks refer to the rank of the lead coefficient. $rank(\Psi)$ is the rank of the lead coefficient matrix in the DGP.
\end{table}

We can notice that the size of the LR test when the true rank of the lead coefficient matrix is 2 is similar to the bivariate case. With a finite variance errors distribution the size of the LR test is around 5\% and it increases to around 9\% when the variance is infinite ($\lambda=1.5$). We can see that the size of the test decreases in the more restrictive CB specification, when the rank of the matrix is 1. For the finite variance cases the size decreases to 91.9\% when $T=500$ and to 93.3\% when $T=1\,000$. The correctly detected model frequency decreases further to 86\% in the infinite variance case. Under the alternative of no CB, with finite variance and a sample size of $T=500$, the LR test wrongly detects a bubble (\textit{2 vs 3}) in 30.5\% of the cases, however this frequency decreases to 6.8\% when the sample size increases to 1\,000. Hence, it seems that with a smaller sample size and the finite variance of the errors distribution, estimating 8 coefficients in the lead matrix instead of 9 in the unrestricted model still provide a good enough fit to not be rejected by the test. The power of the test for all other model specification is above 99.7\%.\footnote{Something to take into account is that under the null of no CB, in the estimations of the restricted models with coefficient matrix of rank 2 or 1, the likelihood function might not have reached the global maximum due to the starting values issue. This could imply an overestimation of the frequencies displayed in the \textit{2 vs 3} and \textit{1 vs 3} when the true rank is 3.} \\

When it comes to model selection using information criteria, BIC outperforms AIC in each of the settings to detect common bubbles. BIC correctly select a model with CB in more than 97.2\% of the cases across all model specifications and the frequencies increase with the sample size and the amount of restricted coefficients. Indeed, it correctly selects a restricted model with a coefficient matrix of rank 1 in at least 99.8\% of the cases. Whereas AIC selects the correct restricted model in less than 88.3\% and the frequency decreases with the sample size, the variance of the errors and when the rank of the restricted matrix is closer to full rank. Hence for the infinite variance case with a sample size $T=500$, its frequency of correctly selected CB model is around 77.5\% for each of the CB specification.  Under the alternative of no CB, we observe the same pattern as for the LR test. In the finite variance case, both information criteria over select a restricted model with a matrix of rank 2. For a sample size of 500, BIC selects the restricted model in 51.9\% of the cases, though it decreases to 19.8\% when the sample size increases to 1\,000. AIC on the other hand only selects the restricted model in 14.5\% with $T=500$ and it even decreases to 3\% with $T=1\,000$. For all other model specification both IC select the correct model in at least 99.4\% of the cases.  \\

Overall, the size of the LR test seems to converge to 5\% in the finite variance cases when the sample size increases. In the infinite variance cases, the size is around 5 percentage points lower and seems to be less affected by the sample size. The power of the test is above 93\% in all model specifications except with $\lambda=3$ and $T=500$, with a restricted model that has only 1 coefficient less to estimate than the unrestricted model (\textit{2 vs 3}). For the model selection using information criteria, BIC overall outperforms AIC in correctly detecting a CB, but also tends to detect a CB more often than AIC when there is none in the \textit{2 vs 3} case with  $\lambda=3$.\footnote{Note that Hannan-Quin information criterion $HQC= 2Kln(ln(T))-2\,\text{ln}(\hat{L})$ performs exactly in between BIC and AIC both under the null and under the alternative. We thus omit it to save space but results are available upon request.}

\section{Common bubbles in commodity indices?}\label{VMAR-sec:empirical}

\begin{figure}[h!]
\begin{subfigure}{\textwidth}
  \centering
    \caption{Indices in levels}
  \includegraphics[width=0.8\linewidth]{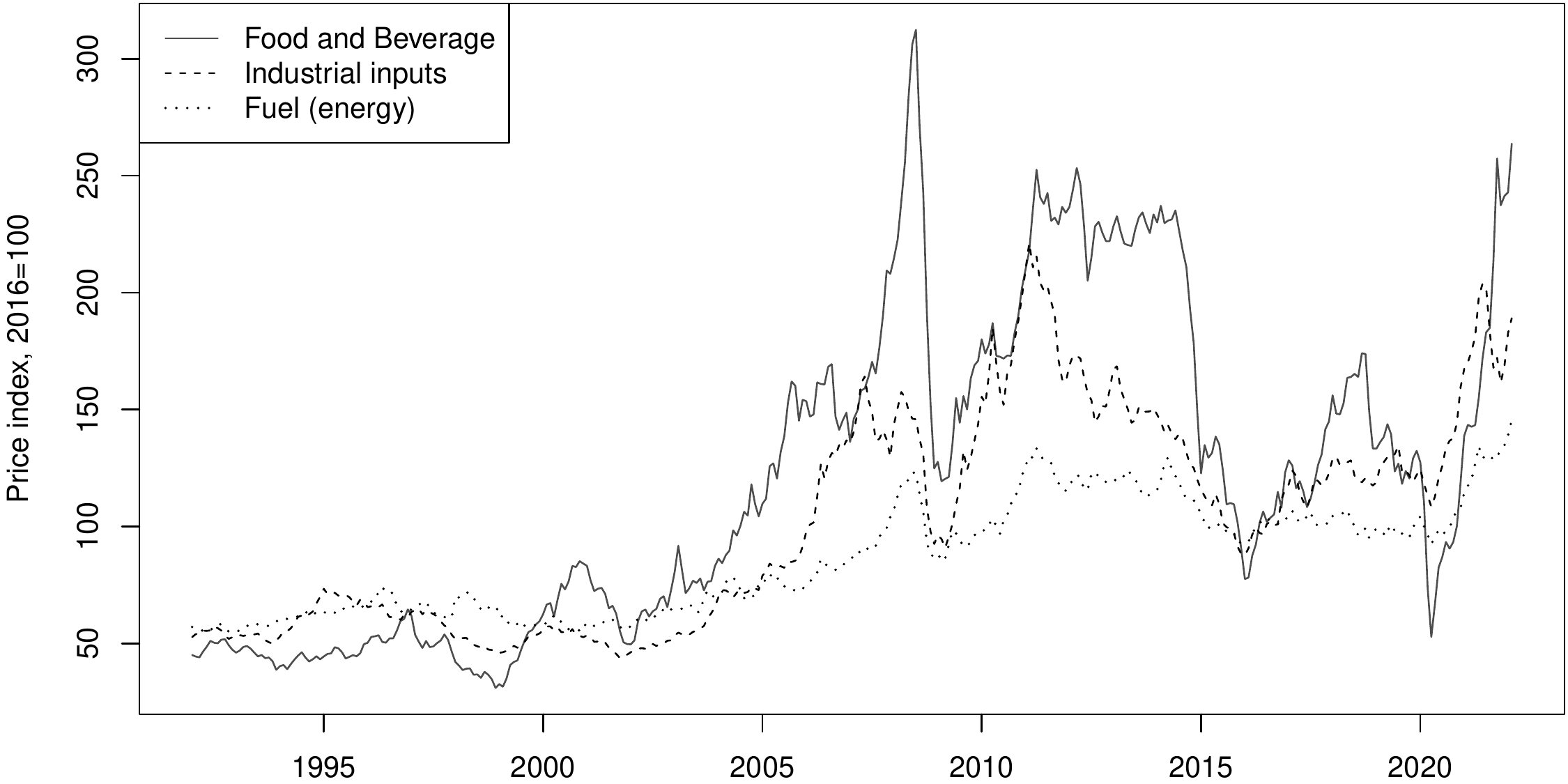}
\end{subfigure}\\
\begin{subfigure}{\textwidth}
  \centering
  \vspace{0.5cm}
    \caption{HP-detrended levels}
  \includegraphics[width=0.8\linewidth]{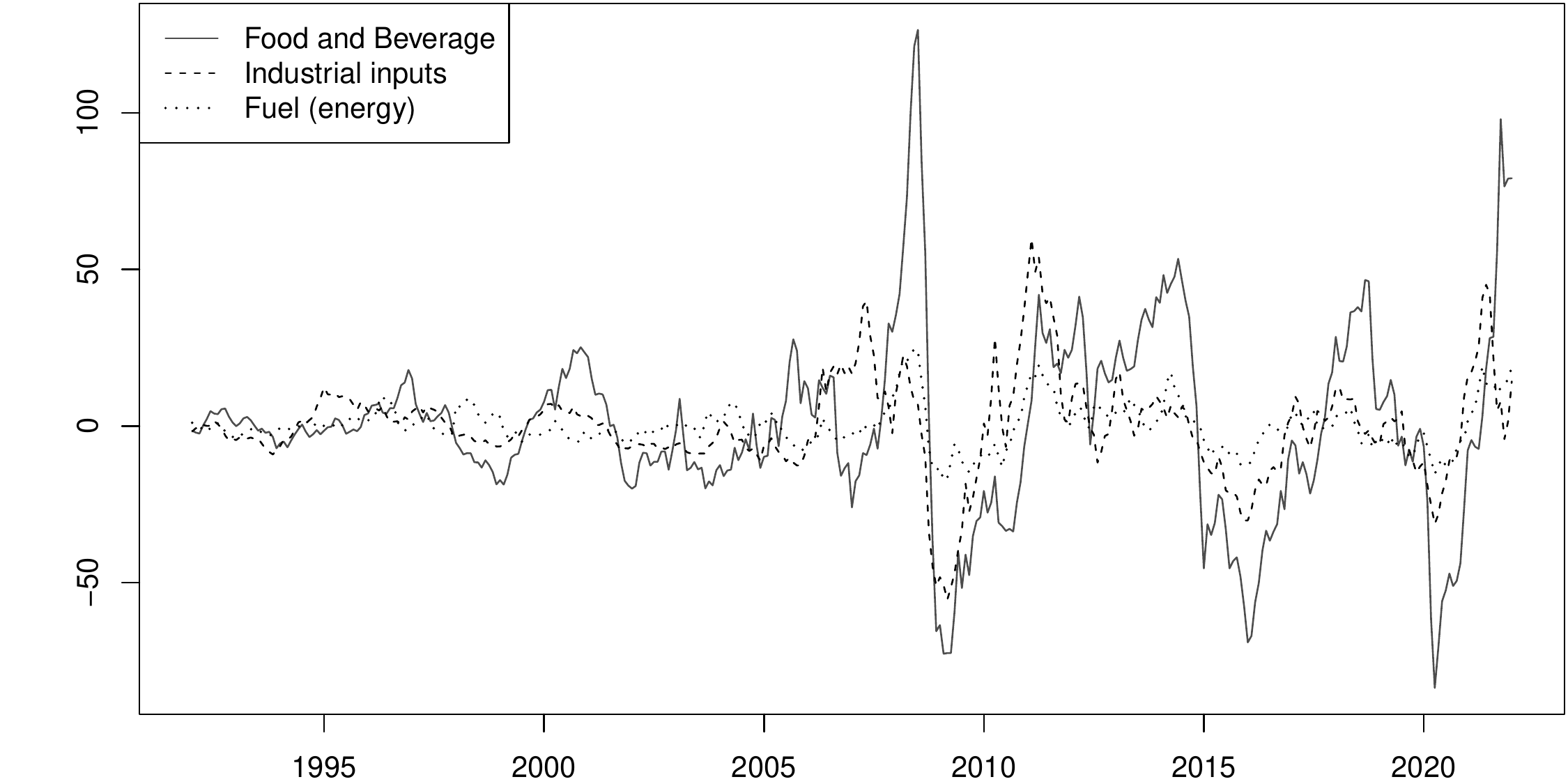}
\end{subfigure}
\caption{Price indices in levels}
\label{VMAR-fig:price_indices_levels}
\end{figure}

\begin{figure}[h!]
\begin{subfigure}{\textwidth}
  \centering
    \caption{Indices in logs}
  \includegraphics[width=0.8\linewidth]{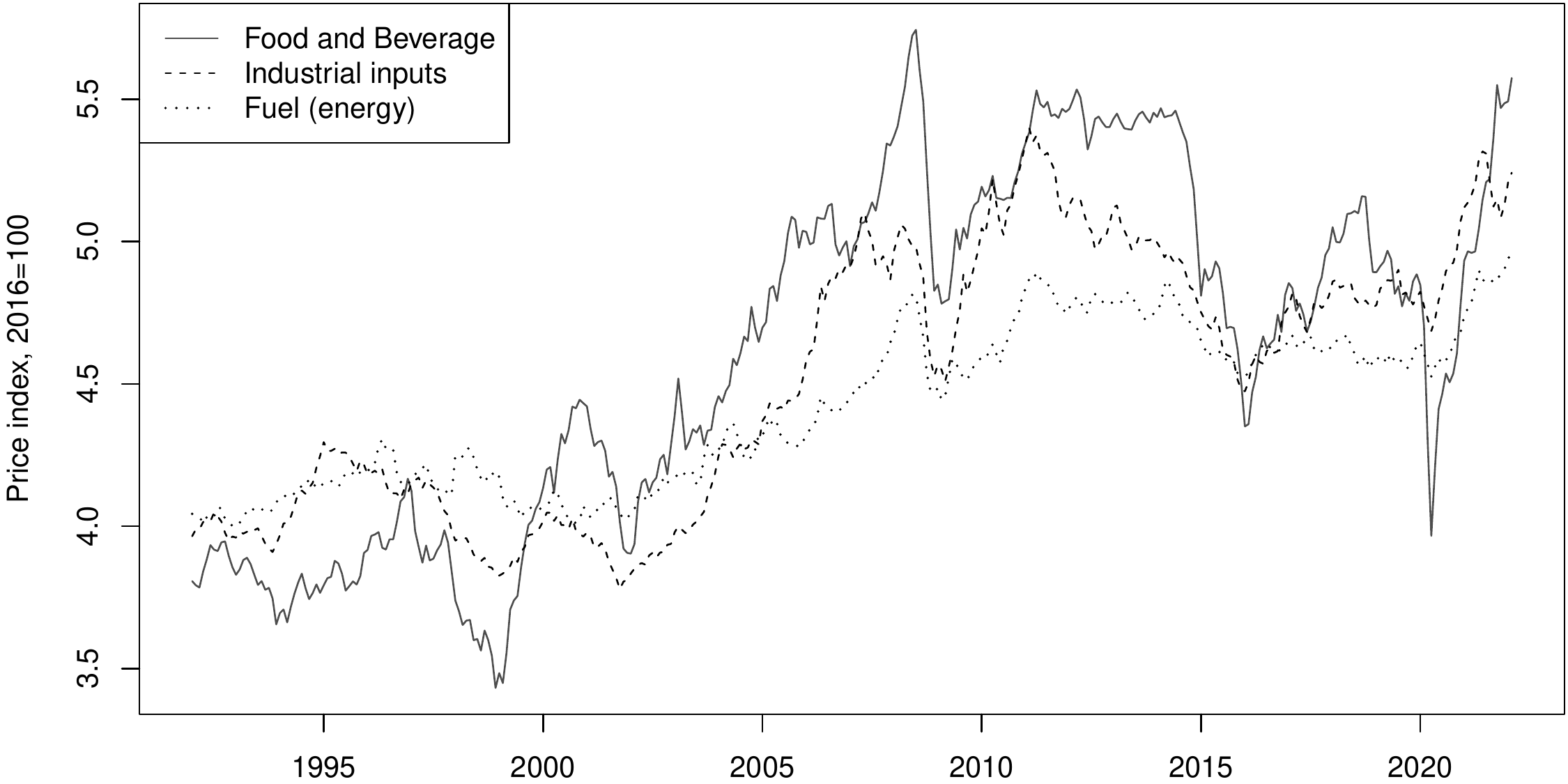}
\end{subfigure}\\
\begin{subfigure}{\textwidth}
  \centering
    \vspace{0.5cm}
    \caption{HP-detrended logs}
  \includegraphics[width=0.8\linewidth]{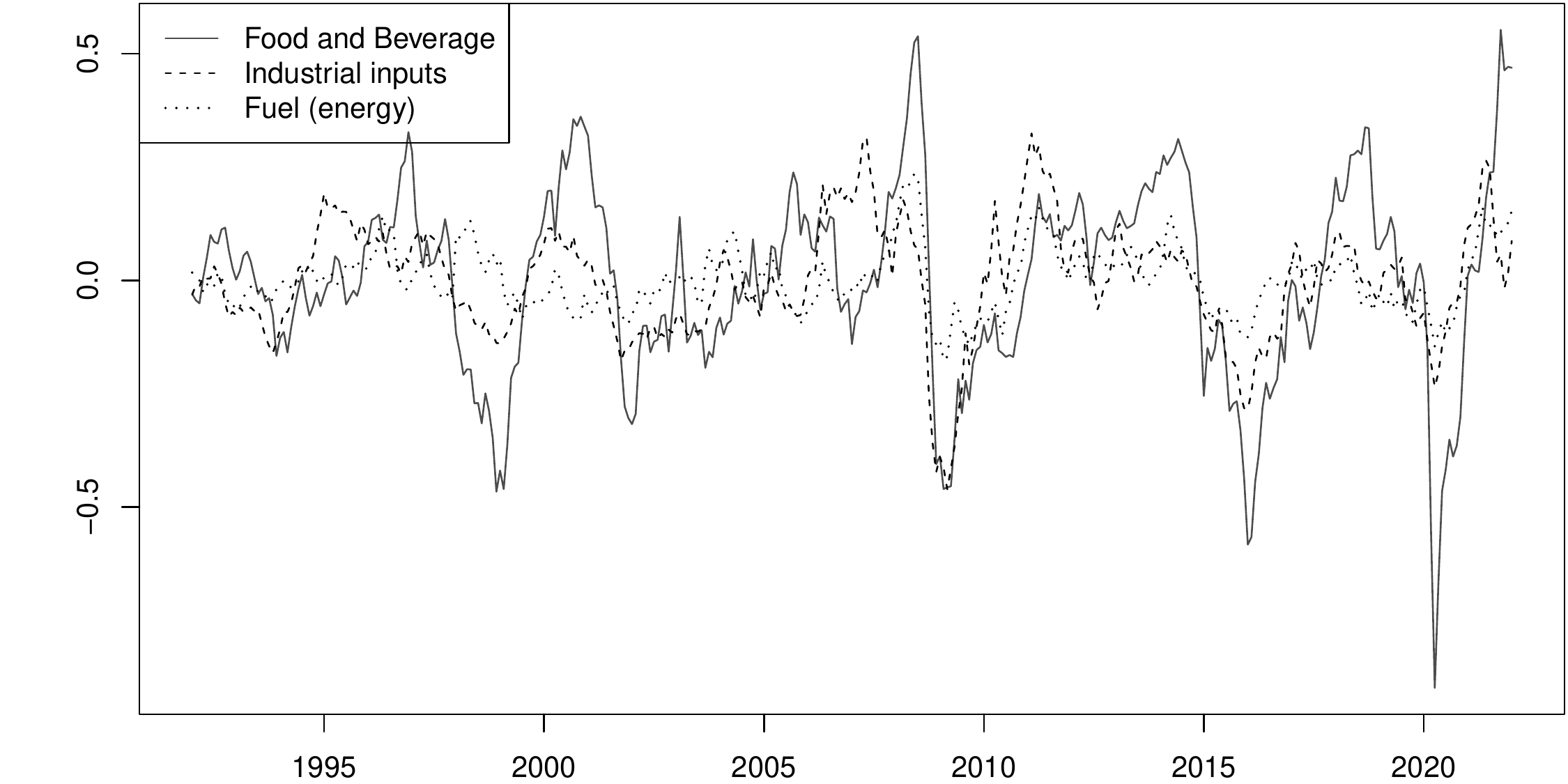}
\end{subfigure}
\caption{Price indices in logs}
\label{VMAR-fig:price_indices_logs}
\end{figure}

We illustrate our strategies to test for common bubbles in mixed causal-noncausal processes on three commodity price indices: food and beverage, industrial inputs\footnote{Includes agricultural raw materials which includes timber, cotton, wool, rubber and hides.} and fuel (energy)\footnote{Includes crude oil, natural gas, coal and propane.}. The sample of 362 data points ranges from January 1992 to January 2022.\footnote{All the data is retrieved from the IMF database. They are price indices with base year 2016.} We can see from graphs (a) of Figures \ref{VMAR-fig:price_indices_levels} and \ref{VMAR-fig:price_indices_logs}, which respectively shows the series in levels and logs, that the indices seem to follow similar trends. Long-lasting increases and crashes often happen at the same time. This could potentially suggest the presence of common bubbles between the series. Following the work of \citet{voisin2021oil}, we detrend all series using the Hodrick-Prescott filter (hereafter HP filter). Although this approach to get stationary time series has been strongly criticized, in particular for the investigation of business cycles, \citet{voisin2021oil} show that it is a convenient strategy to preserve the bubble features. They also show in a Monte Carlo simulation that this is the filter that preserves the best the identification of the MAR($r,s$) model. \citet{giancaterini2022climate} end up to the same conclusions, reinforce the same conclusion using analytical arguments. \\

Detrended series are displayed on graphs (b) of the two Figures. It can indeed be seen that the dynamics inherent to mixed causal-noncausal processes mentioned above are preserved. The crashes occurring during the financial crisis of 2007 and the COVID-19 pandemic in 2020, while being of different magnitude, happened at the same time on all three series. Furthermore, long lasting increases such as the one before the financial crash, the recovery around 2009 or after 2020 are also present in all three index prices. \\

We first analyze the series individually. We estimate pseudo causal autoregressive models to identify the order of autocorrelation in each of the detrended series (both in levels and logs). All models that we identify using BIC end up to be  AR(2) processes. The normality of the errors is rejected for all series: values of the Jarque-Bera statistics range between 48 and 253 for the 6 series. The next step is to identify MAR($\textit{r},\textit{s}$) models for all \textit{r} and \textit{s} subject to the constraint $p=r+s=2$, namely MAR(2,0), MAR(1,1) or MAR(0,2). Based on the maximum likelihood estimator with Student's \textit{t}-distributed error term, the best fitting model for all six series is a MAR(1,1) model.\\

The estimated models are shown in Table \ref{VMAR-tab:commodities_estimated_coeff_univariate}.\footnote{We use different starting values in the estimation to account for the bimodality of the coefficients \citep[see][for more details]{bec2020mixed}.} For comparison purposes with the trivariate case shown later, we display both the coefficients estimated from the multiplicative from
\begin{equation}\label{VMAR-eq:multiplicative_emp}
    (1-\phi L)(1-\psi L^{-1})y_t=\varepsilon_t, \quad\quad \text{with} \quad \varepsilon_t\sim t(\lambda),
\end{equation}
but also the coefficients $b_1$ and $b_2$ of the expanded form obtained after estimations, 
\begin{equation}\label{VMAR-eq:expanded_emp}
    \begin{split}
        y_t&=\frac{\phi}{1+\phi\psi}y_{t-1}+\frac{\psi}{1+\phi\psi}y_{t+1}+\varepsilon^*_t \\
        &=b_1y_{t-1}+b_2y_{t+1}+\varepsilon^*_t.
    \end{split}
\end{equation}

It emerges that ``food and beverage" as well as ``industrial inputs" are both mostly forward looking with lead coefficients close to 0.85 and lag coefficients around 0.4. On the opposite, the ``fuel index" appears more backward looking with coefficients inverted. Except for the level of industrial inputs, all models have error terms with finite variance, and as expected, one obtains lower variance for the logs of the series. The similar dynamics between food and beverages and industrial inputs could indicate commonalities. The same conclusions can be drawn from the rescaled expanded coefficients. \\

\begin{table}[h!]
\caption{Estimated coefficients on univariate MAR(1,1) models}
\label{VMAR-tab:commodities_estimated_coeff_univariate}\centering
\begin{threeparttable}
        \begin{tabular}{ll ccc cc cc}
            \hline
              {\multirow{3}{*}{Variable}} &  & \multicolumn{7}{c}{Estimated coefficients} \\ \cline{3-9}
                                        & & \multicolumn{3}{c}{Multiplicative} &&& \multicolumn{2}{c}{Expanded} \\
            \multicolumn{1}{c}{}                          &  & $\phi$   &     $\psi$   &     $\lambda$ && &$b_1$&$b_2$ \\ \hline
            Food \& Beverage &   & 0.38 &   0.85 &   3.70 & &&0.29 &0.64\\
            log(Food \& Beverage) &   & 0.34 &   0.86 &   5.47 & &&0.26 &0.67\\
             &   &  &  &  &  &  & \\
            Industrial inputs &   & 0.43 &   0.87 &   1.66&& &0.31 &0.63 \\
            log(Industrial inputs) &   & 0.42 &   0.89 &   4.62& &&0.31&0.65 \\
             &   &  &  &   & &&\\
            Fuel (energy) &   & 0.87 &   0.44 & 2.20& &&0.63&0.32 \\
            log(Fuel) &   & 0.83 &  0.48 &  4.95&& &0.59&0.34 \\
            \hline
        \end{tabular}%
    \begin{tablenotes}
      \footnotesize
      \item The coefficients in the multiplicative form are the estimated coefficients from equation \eqref{VMAR-eq:multiplicative_emp}. The expanded coefficients are the ones obtained after expanding the multiplicative from like in \eqref{VMAR-eq:expanded_emp}. 
    \end{tablenotes}
  \end{threeparttable}
\end{table}

For the multivariate investigations, we analyze both bivariate and trivariate systems. Similarly to the univariate estimation, the strategy consists in first estimating the pseudo lag order \textit{p} using a standard VAR($p$) for the six bivariate combinations (three in levels and three in logs) and the two trivariate models. Using BIC, all VARs are identified as VAR(2). There are starting values issues when estimating VMARs by maximum likelihood, meaning that we often reach local maxima. To avoid this, we used a large range of starting values to estimate VMAR(1,1) with multivariate Student's \textit{t}-distributed errors and we keep the estimated model with the highest likelihood value.\footnote{We fixed the starting values for the correlation matrix $\Sigma $ and the degrees of freedom $\lambda $ and performed 100 MLEs based on random lead and lag coefficient matrices fulfilling stationary conditions.}\\

The estimated models are shown below in Table \ref{VMAR-tab:commodities_estimated_coeff_multivariate}. We employ representation \eqref{VMAR-eq:VMAR_multiplicative_psi_phi} for the estimation but the coefficients displayed are those of the additive form \eqref{VMAR-eq:additive}, which are independent of the representation used for the estimations in the following form,
\begin{equation*}
    Y_{t}=B_{1}Y_{t-1}+B_{-1}Y_{t+1}+\eta _{t},
\end{equation*}%
where $\eta _{t}$ follows a multivariate Student's \textit{t}-distribution with $\lambda $ degrees of freedom and correlation matrix $\Omega $. \\

\begin{table}[h!]
\caption{Estimated coefficients on the multivariate VMAR(1,1) models}
\label{VMAR-tab:commodities_estimated_coeff_multivariate}\centering
\resizebox{\textwidth}{!}{%
\begin{threeparttable}
        \begin{tabular}{ccc ccc  ccc c}
            \hline
            \multicolumn{3}{c}{$B_1$} & \multicolumn{3}{c}{$B_{-1}$} & \multicolumn{3}{c}{$\Omega$} & $\lambda$ \\ \hline \vspace{-0.1cm} \\
            \multicolumn{10}{l}{Food and Indus} \\ \vspace{-0.4cm} \\ 
            \multicolumn{3}{c}{$\begin{bmatrix}0.28 & 0.01 \\ 0.26 & 0.27\end{bmatrix}$}&
            \multicolumn{3}{c}{$\begin{bmatrix}0.65 & -0.02 \\ -0.11 & 0.65\end{bmatrix}$}& 
            \multicolumn{3}{c}{$\begin{bmatrix}1.32 & 0.16 \\ 0.16 & 3.35\end{bmatrix}$} & 2.49 \\
            \vspace{-0.1cm} \\
            \multicolumn{10}{l}{Food and Fuel} \\ 
            \vspace{-0.4cm} \\
            \multicolumn{3}{c}{$\begin{bmatrix}0.35 & 0.05 \\ 0.47 & 0.52\end{bmatrix}$}&
            \multicolumn{3}{c}{$\begin{bmatrix}0.55 & -0.04 \\ -0.40 & 0.40\end{bmatrix}$}& 
            \multicolumn{3}{c}{$\begin{bmatrix}1.42 & 0.87 \\ 0.87 & 12.90\end{bmatrix}$} & 3.01 \\
            \vspace{-0.1cm} \\
            \multicolumn{10}{l}{Indus and Fuel} \\ 
            \vspace{-0.4cm} \\
            \multicolumn{3}{c}{$\begin{bmatrix}0.29 & 0.01 \\ -0.11 & 0.47\end{bmatrix}$}&
            \multicolumn{3}{c}{$\begin{bmatrix}0.63 & 0.03 \\ 0.09 & 0.48\end{bmatrix}$}& 
            \multicolumn{3}{c}{$\begin{bmatrix}2.22 & 1.50 \\ 1.50 & 7.17\end{bmatrix}$} & 1.67 \\
            \vspace{-0.1cm} \\
            \multicolumn{10}{l}{Food, Indus and Fuel} \\ 
            \vspace{-0.4cm} \\
            \multicolumn{3}{c}{$\begin{bmatrix}0.27 & 0.01 & 0.03 \\ 0.27 & 0.25 & 0.02 \\ 0.30 & -0.10 & 0.56\end{bmatrix}$}&
            \multicolumn{3}{c}{$\begin{bmatrix}0.64 & -0.02 & -0.02 \\ -0.16 & 0.66 & -0.01 \\ -0.27 & 0.12 & 0.37\end{bmatrix}$}& 
            \multicolumn{3}{c}{$\begin{bmatrix}1.34 & 0.12 & 0.55 \\ 0.12 & 3.29 & 2.20 \\ 0.55 & 2.20 & 10.02\end{bmatrix}$} & 2.28 \\ \\

            \hline
            \multicolumn{3}{c}{$B_1$} & \multicolumn{3}{c}{$B_{-1}$} & \multicolumn{3}{c}{$10^3\Omega$} & $\lambda$ \\ \hline \vspace{-0.1cm} \\
            \multicolumn{10}{l}{Food and Indus} \\ \vspace{-0.4cm} \\ 
            \multicolumn{3}{c}{$\begin{bmatrix}0.25 & 0.01 \\ 0.22 & 0.24\end{bmatrix}$}&
            \multicolumn{3}{c}{$\begin{bmatrix}0.69 & -0.03 \\-0.12 & 0.70\end{bmatrix}$}& 
            \multicolumn{3}{c}{$\begin{bmatrix}0.27 & 0.03 \\ 0.03 & 0.46\end{bmatrix}$} & 6.30 \\
            \vspace{-0.1cm} \\
            \multicolumn{10}{l}{Food and Fuel} \\ 
            \vspace{-0.4cm} \\
            \multicolumn{3}{c}{$\begin{bmatrix}0.25 & 0.02 \\ 0.16 & 0.38\end{bmatrix}$}&
            \multicolumn{3}{c}{$\begin{bmatrix}0.67 & -0.02 \\ -0.14 & 0.55\end{bmatrix}$}& 
            \multicolumn{3}{c}{$\begin{bmatrix}0.25 & 0.06 \\ 0.06 & 1.21\end{bmatrix}$} & 5.23 \\
            \vspace{-0.1cm} \\
            \multicolumn{10}{l}{Indus and Fuel} \\ 
            \vspace{-0.4cm} \\
            \multicolumn{3}{c}{$\begin{bmatrix}0.26 & 0.04 \\ -0.09 & 0.56\end{bmatrix}$}&
            \multicolumn{3}{c}{$\begin{bmatrix}0.67 & -0.01 \\ 0.09 & 0.37\end{bmatrix}$}& 
            \multicolumn{3}{c}{$\begin{bmatrix}0.42 & 0.24 \\ 0.24 & 1.19\end{bmatrix}$} & 4.77 \\
            \vspace{-0.1cm} \\
            \multicolumn{10}{l}{Food, Indus and Fuel} \\ 
            \vspace{-0.4cm} \\
            \multicolumn{3}{c}{$\begin{bmatrix}0.88 & -0.17 & -0.02 \\ -0.04 & 0.27 & 0.07 \\-0.04 & 0.06 & 0.58\end{bmatrix}$}&
            \multicolumn{3}{c}{$\begin{bmatrix}0.21 & 0.15 & 0.00 \\ 0.13 & 0.76 & -0.08 \\ 0.02 & 0.05 & 0.33\end{bmatrix}$}& 
            \multicolumn{3}{c}{$\begin{bmatrix}0.32 & 0.02 & 0.07 \\ 0.02 & 0.51 & 0.26 \\ 0.07 & 0.26 & 1.35\end{bmatrix}$} & 6.15 \\ \\
            \hline
        \end{tabular}%
  \end{threeparttable}
  }
\end{table}

Comparing to the expanded coefficients $b_1$ and $b_2$ of the univariate models in \ref{VMAR-tab:commodities_estimated_coeff_univariate}, the directions and magnitudes of the dynamics have been preserved in the multivariate models estimations. From the off-diagonal coefficients of the bivariate models, we notice that `Food' is impacting both `Indus' and `Fuel' with the lag and the lead, with coefficients magnitude between 0.11 and 0.47 for the levels. However in the other direction, the magnitude of the coefficients does not exceed 0.05 for the lag of `Fuel' on `Food'. `Fuel' slightly impacts `Indus' with coefficients of magnitude around 0.1. These dynamics can also be observed in the trivariate model. \\

To perform the common bubble tests we estimated VMAR models with restrictions on the lead coefficients matrix as shown in \eqref{VMAR-eq:VMAR_restricted}.\footnote{We also used 100 combinations of starting values to make sure we obtain the best fitting models.} In the trivariate settings the LR test and the information criteria compares the unrestricted model where the lead matrix has full rank with both CB specifications, namely imposing rank 2 or rank 1 to the lead coefficient matrix.\\

The results are shown in Table \ref{VMAR-tab:commodities_CB_tests}. The LR column displays the LR test statistic and the IC columns are the difference in the IC values of the restricted and the unrestricted models. Looking at the LR tests, the null hypothesis of a common bubble in the bivariate and trivariate models is rejected for all combination of variables at a confidence level of 95\%. All information criteria also indicate a better fit for the models without commonalities since all values are positive. Even for the trivariate cases \textit{2 vs 3}, no bubble is detected even though in the simulations exercise, the test and information criteria over-detected a CB for such sample size and degrees of freedom. Hence, while the series seem to follow similar pattern in the locally explosive episodes throughout the time period, we do not find significant indication of commonalities in their forward looking components. \\

\begin{table}[h!]
\caption{Common bubble detection on multivariate combinations of the variables}
\label{VMAR-tab:commodities_CB_tests}\centering
\begin{threeparttable}
\begin{tabular}{ccccccccc}
\hline
\multicolumn{3}{l}{Levels}                                                                                                                                      &                      & \multicolumn{5}{c}{ }                                                                           \\  
Food                                                & Indus                                               & Fuel                                                & Rank test            & LR                   &                      & BIC                  &                      & AIC                  \\ \hline
$\blacksquare$                                      & $\blacksquare$                                      &                                                     & 1 vs 2               & 25.93 && 20.04 && 23.93                 \\
$\blacksquare$                                      &                                                     & $\blacksquare$                                      & 1 vs 2               & 59.96 && 54.07 && 57.96                 \\
                                                    & $\blacksquare$                                      & $\blacksquare$                                      & 1 vs 2               & 70.49 && 64.59 && 68.49                \\
                                  &                                   &                                   &    &    &    &    &    &    \\
\multirow{2}{*}{$\blacksquare$}                     & \multirow{2}{*}{$\blacksquare$}                     & \multirow{2}{*}{$\blacksquare$}                     & 2 vs 3               & 16.26 && 10.37 && 14.26                      \\
                                                    &                                                     &                                                     & 1 vs 3               & 88.12 && 64.55 && 80.12                      \\
                                                    &                                                     &                                                     &                      &                      &                      &                      &                      &                      \\
\multicolumn{3}{l}{Logs}                                                                                                                                        &    &    &    &    &    &    \\
Food                                                & Indus                                               & Fuel                                                & Rank test            & LR                   &                      & BIC                  &                      & AIC                  \\ \hline
$\blacksquare$                                      & $\blacksquare$                                      &                                                     & 1 vs 2               & 16.04 && 10.15 && 14.04                 \\
$\blacksquare$                                      &                                                     & $\blacksquare$                                      & 1 vs 2               & 34.36 && 28.47 && 32.36                \\
                                                    & $\blacksquare$                                      & $\blacksquare$                                      & 1 vs 2               & 46.05 && 40.16 && 44.05                \\
                                  &                                   &                                   &    &    &    &    &    &    \\
  {\multirow{2}{*}{$\blacksquare$}} &   {\multirow{2}{*}{$\blacksquare$}} &   {\multirow{2}{*}{$\blacksquare$}} & 2 vs 3               & 15.81 &&  9.92 && 13.81                     \\
                                  &                                   &                                   & 1 vs 3               & 75.01 && 51.44 && 67.01                      \\\hline
\end{tabular}%
        \begin{tablenotes}
          \footnotesize
          \item LR is the likelihood ratio test statistic. For the bivariate models the critical value of the LR test at 95\% confidence level is 3.41. For the trivariate models, the critical values are 3.841 and 9.488 for \textit{2 vs 3 }and \textit{1 vs 3} respectively. The column BIC and AIC show the difference between the restricted and unrestricted information criteria. 
        \end{tablenotes}
    \end{threeparttable}
\end{table}

\section{Conclusion}\label{VMAR-sec:conclusion}
This paper proposes methods to investigate whether the bubble patterns observed in individual series are common to various series. We detect the non-linear dynamics using the recent mixed causal and noncausal models. The lead component of the model allows to capture, for instance, locally explosive episodes in a parsimonious and strictly stationary setting. We hence employ multivariate mixed causal-noncausal models and apply restrictions to the lead coefficients matrices to test for the presence of commonalities in the forward looking components of the series. We propose a likelihood ratio (LR) test to test for the presence of a common bubble. In a simulation study, we investigate the accuracy of the common bubbles detection using the LR test as well as by model selection using information criteria. Then, implementing our approach on three commodity prices we do not find evidence of commonalities despite the similarities between the series. Our definition of common bubbles requires that all noncausal matrices span the same left null space. A natural extension to our approach would be to relax that hypothesis to investigate non synchronous common bubbles, allowing for some adjustment delays along the lines of \citet{cubadda2001non}.

\newpage
\section*{Acknowledgments}
Elisa Voisin gratefully acknowledges the University of Rome Tor Vergata for organizing a 3-month research visit, during which this paper was partially written.

\bibliography{references.bib}

\end{document}